\documentclass[a4paper,11pt]{article}

\usepackage{jcappub} 

\renewcommand{\thesection}{\arabic{section}}

\usepackage{hyperref}
\usepackage{amsmath}
\usepackage{amssymb}
\usepackage{subfigure}
\usepackage{graphicx}
\usepackage{color}
\usepackage[usenames,dvipsnames,svgnames,table]{xcolor}

\renewcommand{\thetable}{\arabic{table}}
\numberwithin{equation}{section}

\title{\boldmath Color flavor locked strange stars in 4D Einstein-Gauss-Bonnet gravity }

\author[a]{Ayan Banerjee\note{Corresponding author}}
\author[b,c]{and Ksh. Newton Singh}

\affiliation[a]{Astrophysics and Cosmology Research Unit, University of KwaZulu Natal, Private Bag X54001, Durban 4000,
South Africa,}
\affiliation[b]{Department of Physics, National Defence Academy, Khadakwasla, Pune-411023, India,}
\affiliation[c]{Department of Mathematics, Jadavpur University, Kolkata-700032, India,}

\emailAdd{ayan\_7575@yahoo.co.in}
\emailAdd{ntnphy@gmail.com}

\abstract{This work is devoted on the recently introduced Einstein-Gauss-Bonnet gravity in four dimensions.
The theory can bypass the Lovelock's theorem and avoids Ostrogradsky instability. The integrated part of this theory is the GB term gives rise to a non-trivial contribution to the gravitational dynamics in the limit $D \to 4$. Our main interest is to explore a class of static and spherically symmetric compact objects made of strange matter in the color flavor locked (CFL) phase, in which all three quark flavors participate to pairing symmetrically. It is therefore natural to ask for compact objects in 4$D$ EGB gravity since, compact stars signatures may be the test-bed to compare new models to the Einstein gravity; on the other hand, this could shed new light on the possible presence of CFL phase in compact stars. This feature may be relevant in view of recent observations claiming the existence of stable compact \emph{quark} stars (hybrid neutron stars or strange stars). }   


\begin{document}
\maketitle
\flushbottom

\section{Introduction}\label{intro:Sec}

Low-energy effective string theories arising from various compactification schemes. The main features of low-energy string theory which appears as an effective model of gravity in higher dimensions that involve higher order curvature terms in the action. In fact, this theory has been proposed with the hope that higher order corrections to Einstein's GR might solve the singularity problem of black holes and early universe. Indeed, there exist a lot of works on higher derivative gravity theories ranging from condensed matter, string theory and gravitation, mainly. In this regard, Lovelock theory of gravity, as a natural generalization of Einstein’s general relativity, was proposed by Lovelock \cite{Lovelock,Lovelock:1972vz}. In particular, the equations of motion of Lovelock gravity are of second order with respect to metric, as the case of general relativity. Interestingly, Einstein-Gauss-Bonnet gravity \cite{Lanczos:1938sf} is consider as a special case of Lovelock’s theory of gravitation. The EGB gravity adds an extra term to the standard Einstein-Hilbert action, which is quadratic in the Riemann tensor. It appears naturally in the low energy effective action of heterotic string theory \cite{Wiltshire:1985us,Boulware:1985wk,Wheeler:1986}. \\

Recently, Glavan \& Lin \cite{Glavan:2019inb} has proposed a new covariant modified theory of gravity in 4-dimensional spacetime, namely, `Einstein Gauss-Bonnet gravity' (EGB). In $D$-dimensions, by rescaling the Gauss-Bonnet coupling $\alpha$ by a factor of $\alpha/(D -4)$, and taking the limit $D \to 4$, the Gauss-Bonnet term gives rise to non-trivial dynamics.  By this re-scaling one can bypasses the conclusions of Lovelock’s theorem and avoids Ostrogradsky instability. Though the dimensional regularization of this was considered previously in \cite{Tomozawa:2011gp}. The proposed new or modified version of the theory has attracted researchers for several novel predictions in cosmology and astrophysics, though the validity of this theory is at present under debate and doubts. For example, static spherically symmetric  vacuum black hole is obtained in \cite{Glavan:2019inb}, which differs from the standard vacuum-GR Schwarzschild BH. Related to this other references about rotating and non-rotating  black hole solutions and their physical properties have been discussed 
\cite{Ghosh:2020syx,Konoplya:2020juj,Kumar:2020uyz,Kumar:2020xvu,Zhang:2020sjh,Liu:2020vkh,li04,li05,we03} and the references therein. Moreover, within the same context, the strong/ weak gravitational lensing by black hole  \cite{Islam:2020xmy,Kumar:2020sag,Heydari-Fard:2020sib,Jin:2020emq}, spinning test particle \cite{zh03}, geodesics motion and shadow \cite{Zeng:2020dco}, thermodynamics of 4D EGB AdS black hole \cite{sa03}, Hawking radiation \cite{Zhang:2020qam,Konoplya:2020cbv}, Quasinormal modes \cite{Churilova:2020aca,Mishra:2020gce,ar04}, 
wormhole solution \cite{Jusufi:2020yus,liu20} have been addressed. Additionally, many aspects of this theory have indeed been exploded  \cite{Jusufi:2020qyw,Yang:2020jno,Ma:2020ufk,si20}. \\

Hence, the 4$D$ EGB gravity witnessed significant
attentions that includes finding astrophysical solutions
and investigating their properties. In particular, the
mass-radius relations are obtained for realistic hadronic and for strange quark star EoS \cite{Doneva:2020ped}. Precisely speaking, we are interested to investigate the behaviour of compact star namely neutron/quark stars in regularized 4$D$ EGB gravity. In this article, we investigate the color-flavor-locked (CFL) matter \cite{Alford:1998mk} and its stability related to the  strange quark star.  In particular, the three-flavor quark matter with the particular symmetry is called the CFL matter, which is different from strange quark matter (SQM) and matter without the Bardeen-Cooper-Schrieffer pairing \cite{Alford:1997zt}. Since,  quarks in the cores of neutron stars are likely to be in a paired phase \cite{RAJAGOPAL2001,alford2007}. In addition, pairing affects the spectrum of quasi-particles and can change the transport properties qualitatively. At asymptotically high densities, one  finds quark matter in the CFL phase \cite{Alford:1998mk}.  Indeed, as a strongly interacting matter, the CFL matter is widely accepted to become  `absolutely' stable for sufficiently high densities \cite{Alford2004}.\\

In \cite{Alford2001}, authors have concluded that CFL is more stable than SQM as long as $\mu \gtrsim$  $m_s^2/4 \Delta$,
with $m_s$ being the strange quark mass and $\Delta$ the pairing gap. However,  SQM is more stable than the hadronic phase \cite{Bodmer,Edward}. If the baryon chemical potential is sufficiently large enough, the formation of Copper pairs is favoured due to CFL producing color-superconducting phase, and is believed to be more stable than SQM \cite{Alford:1998mk}. Also, large baryon chemical potential makes the strange quark matter massless as well as CFL, which infers same number density of $u,~d$ and $s$ quarks making the system electrically neutral \cite{Orsariaa}.  It has also been suggested that two-flavor color superconducting (2SC) phase is highly unlikely in the environment of compact objects because the free energy cost in 2SC phase is much higher than the CFL phase \cite{Alford2002}. Furthermore, CFL matter could be adequate candidates to explain  neutron stars or strange stars \cite{Flores:2017hpb,Flores:2010hpb}.\\

Thus, self-bound stars made up of CFL quark matter may  discuss some recent astrophysical observational data that could shed new light on
the possible existence of exotic and/or deconfined phases in some nearby neutron stars (NS). The plan of this paper is as follows: After the introduction in Sec. \ref{intro:Sec}, we quickly review the field equations in the 4$D$ EGB gravity and show that it makes a nontrivial contribution to gravitational dynamics in 4$D$ in Sec. \ref{sec2}.  In Sec. \ref{sec3} we discuss the EoS for CFL strange matter. In Sec. \ref{sec4}, we discuss the numerical procedure used to solve the field equations. In Sec. \ref{sec5}, is devoted to report the general properties of
the spheres in terms of the CFL strange quark matter. We analyzed the energy conditions as well as other
properties of the spheres, such as sound velocity and adiabatic stability. Finally, in Sec. \ref{sec6}, we conclude.

 \section{Basic equations of EGB gravity} \label{sec2}
The field equation in $D$-dimensional EGB theory (we mostly use geometrized units while deriving various equations, which is, $c = G = 1$) is derived from the following action:
\begin{equation}\label{action}
	\mathcal{I}_{G}=\frac{1}{16 \pi}\int d^{D}x\sqrt{-g}\left[ R +\frac{\alpha}{D-4} \mathcal{L}_{\text{GB}} \right]
+\mathcal{S}_{\text{matter}},
\end{equation}
where $R$ is the Ricci scalar which provides the general relativistic part of the action, and  $g$ denotes the determinant of the metric $g_{\mu\nu}$. Since, the Gauss-Bonnet coupling coefficient $\alpha$ is $[length]^2$.  We consider $\alpha \ge 0$ and in the rest of this section. Here, ${\mathcal{L}}_{\text{GB}}$ is the Einstein-Gauss-Bonnet Lagrangian given by
\begin{equation}
\mathcal{L}_{\text{GB}}=R^{\mu\nu\rho\sigma} R_{\mu\nu\rho\sigma}- 4 R^{\mu\nu}R_{\mu\nu}+ R^2\label{GB}.
\end{equation}

Here, $S_{\text{matter}}$ is the action of the standard perfect fluid matter.
Now, varying the action (\ref{action}) with respect to metric $g_{\mu \nu}$, one
obtains the field equations \cite{Ghosh:2020vpc}
\begin{equation}\label{GBeq}
G_{\mu\nu}+\frac{\alpha}{D-4} H_{\mu\nu}= 8 \pi T_{\mu\nu}~,~~~\mbox{where}~~~~~~T_{\mu\nu}= -\frac{2}{\sqrt{-g}}\frac{\delta\left(\sqrt{-g}\mathcal{S}_m\right)}{\delta g^{\mu\nu}},
\end{equation}
with $T_{\mu\nu}$ stands the energy momentum tensor of matter with the following expression
\begin{eqnarray}
&& G_{\mu\nu} = R_{\mu\nu}-\frac{1}{2}R~ g_{\mu\nu},\nonumber\\
&& H_{\mu\nu} = 2\Bigr( R R_{\mu\nu}-2R_{\mu\sigma} {R}{^\sigma}_{\nu} -2 R_{\mu\sigma\nu\rho}{R}^{\sigma\rho} - R_{\mu\sigma\rho\delta}{R}^{\sigma\rho\delta}{_\nu}\Bigl)- \frac{1}{2}~g_{\mu\nu}~\mathcal{L}_{\text{GB}},\label{FieldEq}
\end{eqnarray}
with $R$ the Ricci scalar, $R_{\mu\nu}$ the Ricci tensor, $H_{\mu\nu}$ is the Lancoz tensor and $R_{\mu\sigma\nu\rho}$ the Riemann tensor, respectively.  Note that the GB terms is total derivative  in 4$D$ space-time, and hence  do not contribute to the field equations. However, by re-scaling the coupling
constant as $ \alpha/(D-4)$, it was shown in Ref. \cite{Ghosh:2020vpc} that 
maximally symmetric spacetimes with curvature scale ${\cal K}$, the variation of the Gauss-Bonnet term
was found
\begin{equation}\label{gbc}
\frac{g_{\mu\sigma}}{\sqrt{-g}} \frac{\delta \mathcal{L}_{\text{GB}}}{\delta g_{\nu\sigma}} = \frac{\alpha (D-2) (D-3)}{2(D-1)} {\cal K}^2 \delta_{\mu}^{\nu}.
\end{equation}
An interesting feature of $4 D$ EGB gravity is that Eq. (\ref{gbc}) does not 
vanish in $D=4$, because of the re-scaled coupling constant \cite{Glavan:2019inb}.

In this work, we consider the general static, spherically symmetric $D$-dimensional metric \cite{Mehdizadeh:2015jra} given by the following line element   
\begin{eqnarray}\label{metric}
ds^2_{D}= - e^{2\Phi(r)}dt^2 + e^{2\Lambda(r)}dr^2 + r^{2}d\Omega_{D-2}^2,  
\end{eqnarray} 
where $d\Omega_{D-2}^2$ is the metric on the unit $(D-2)$-dimensional sphere and 
 $\Phi(r)$ and $\Lambda(r)$ are functions of the radial coordinate $r$, respectively. 

\begin{figure}[h]
\centering
\includegraphics[width=0.6\linewidth]{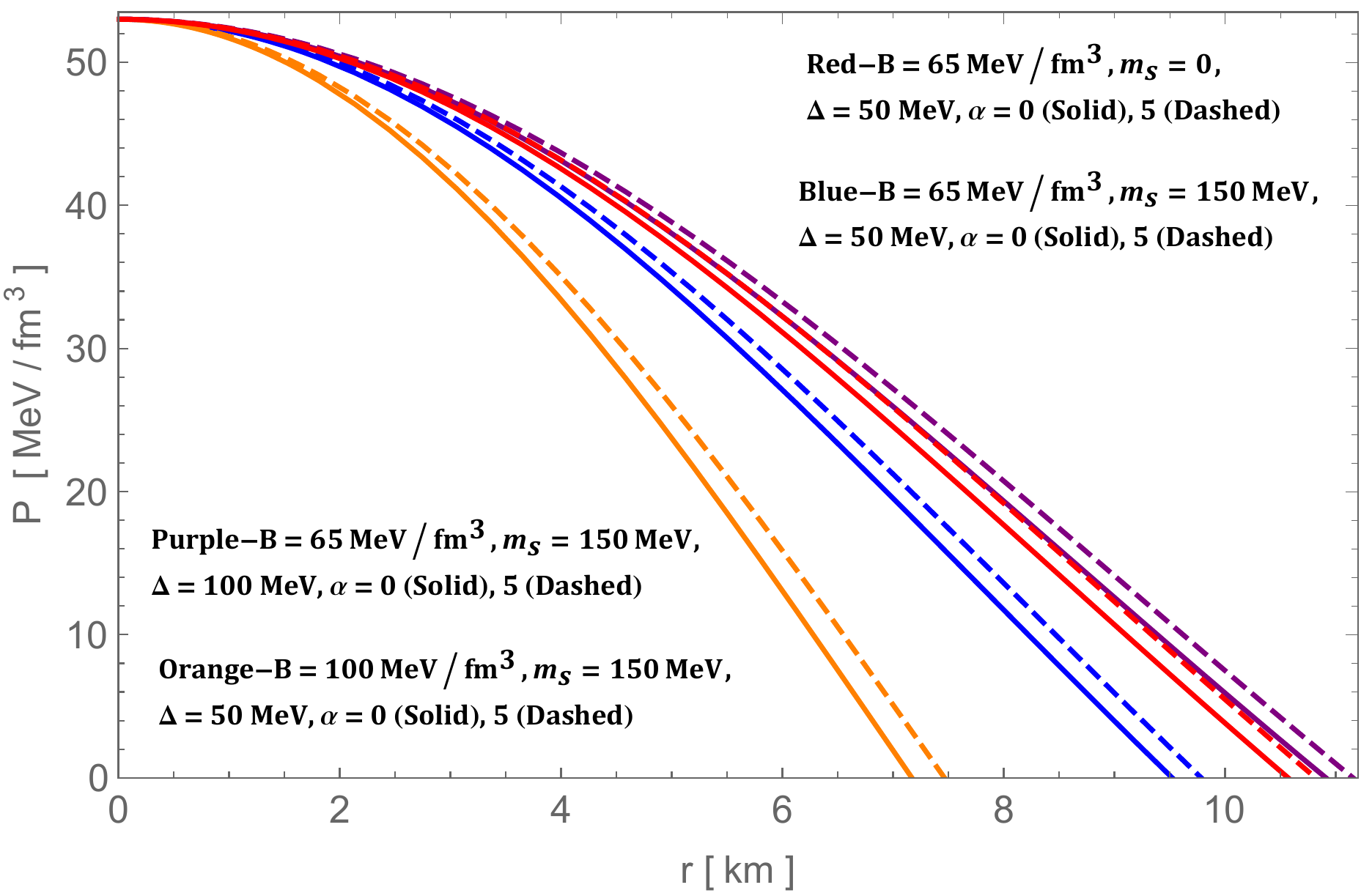}
\caption{Variation of pressure with radius for the EoS of CFL strange quark matter when  $B=65,~100~MeV/fm^3$, $m_s=0,~150~MeV$, $\Delta=50,~100 ~MeV$
and two values of  $\alpha = 0 ~\& ~ 5$.} \label{f1}
\end{figure}

We assume that the interior of star is filled with a perfect fluid matter source with energy-momentum tensor  described by
\begin{eqnarray}
T_{\mu\nu} = (\rho+p)u_{\nu} u_{\nu} + p g_{\nu \nu}, \label{em}
\end{eqnarray}
where $p=p(r)$ is the pressure, $\rho=\rho(r)$ is the energy density of matter, and $u_{\nu}$ is the contravariant $D$-velocity.
On using the metric (\ref{metric}) with stress tensor (\ref{em}), in the limit $D \to 4$,  the $tt$, $rr$ and hydrostatic continuity equations (\ref {GBeq}) read ( dimensionally reduced) :
\begin{eqnarray}\label{DRE1}
&& \frac{2}{r} \frac{d\Lambda}{dr} = e^{2\Lambda} ~ \left[8\pi \rho - \frac{1-e^{-2\Lambda}}{r^2}\left(1-  \frac{\alpha(1-e^{-2\Lambda})}{r^2}\right)\right]\left[1 +  \frac{2\alpha(1-e^{-2\Lambda})}{r^2}\right]^{-1}, \\ 
&& \frac{2}{r} \frac{d\Phi}{dr} = e^{2\Lambda} ~\left[8\pi p + \frac{1-e^{-2\Lambda}}{r^2} \left(1- \frac{\alpha(1-e^{-2\Lambda})}{r^2} \right) \right] \left[1 +  \frac{2\alpha(1-e^{-2\Lambda})}{r^2}\right]^{-1},\label{DRE2} \\
&& \frac{dp}{dr} = - (\rho + p) \frac{d\Phi}{dr}.  \label{DRE3}
\end{eqnarray} 
As usual, the asymptotic flatness imposes $\Phi(\infty)=\Lambda(\infty)=0$ while  the regularity at the center requires $\Lambda(0)=~0$. 

It is advantageous to define  the gravitational mass within the sphere of radius $r$, such that $e^{-2\Lambda} =1-2m(r)/r$.  Now, we are ready to write the Tolman-Oppenheimer-Volkoff (TOV) equations in a form we want to use. So, using (\ref{DRE2}-\ref{DRE3}), we obtain
the modified TOV as
\begin{equation}
{dp \over dr} = -{\rho(r) m(r) \over r^2}\left[1+{p(r) \over \rho(r)}\right]\left[1+{4\pi r^3 p(r) \over m(r)}-{2\alpha m(r) \over r^3}\right]\left[1+{4\alpha m(r) \over r^3}\right]^{-1} \left[1-{2m(r) \over r}\right]^{-1}. \label{e2.11}
\end{equation}
If we take the $\alpha \to 0$ limit, this equation reduces to the standard TOV equation of GR. Replacing the last equality in Eq. (\ref{DRE1}), we obtain  the gravitational mass:
\begin{equation}
m'(r)=\frac{6 \alpha  m(r)^2+4 \pi  r^6 \rho (r)}{4 \alpha  r m(r)+r^4}, \label{e2.12}
\end{equation}
using the initial condition $m(0)=0$. Finally, if we assume an incompressible fluid, i.e., $\rho(r)=const$, so the energy density is constant along the whole star, one gets
\begin{equation}\label{mass}
m(r)=\frac{r^3} {4 \alpha } \left(-1\pm \sqrt{\frac{3+ 32 \pi  \alpha \rho_c }{3}  }\right),
\end{equation}
where $\rho_c$ is constant.  The $\pm$ sign in Eq. (\ref{mass}) refers to two different branches of
solution. In our case, we shall restrict our discussion
to the positive branch of solution as the expression always gives positive solution for $\alpha>0$. For more details discussion we refer our reader Ref. \cite{Doneva:2020ped}.

So, the final three Eqs. (\ref{DRE1})-(\ref{DRE2}) and (\ref{e2.11}) can be solved numerically for a given EoS $p=p(\rho)$. Here, we are interested to consider color flavor locked strange matter to solve the field equations.

\begin{figure}[h]
\centering
\includegraphics[width=0.6\linewidth]{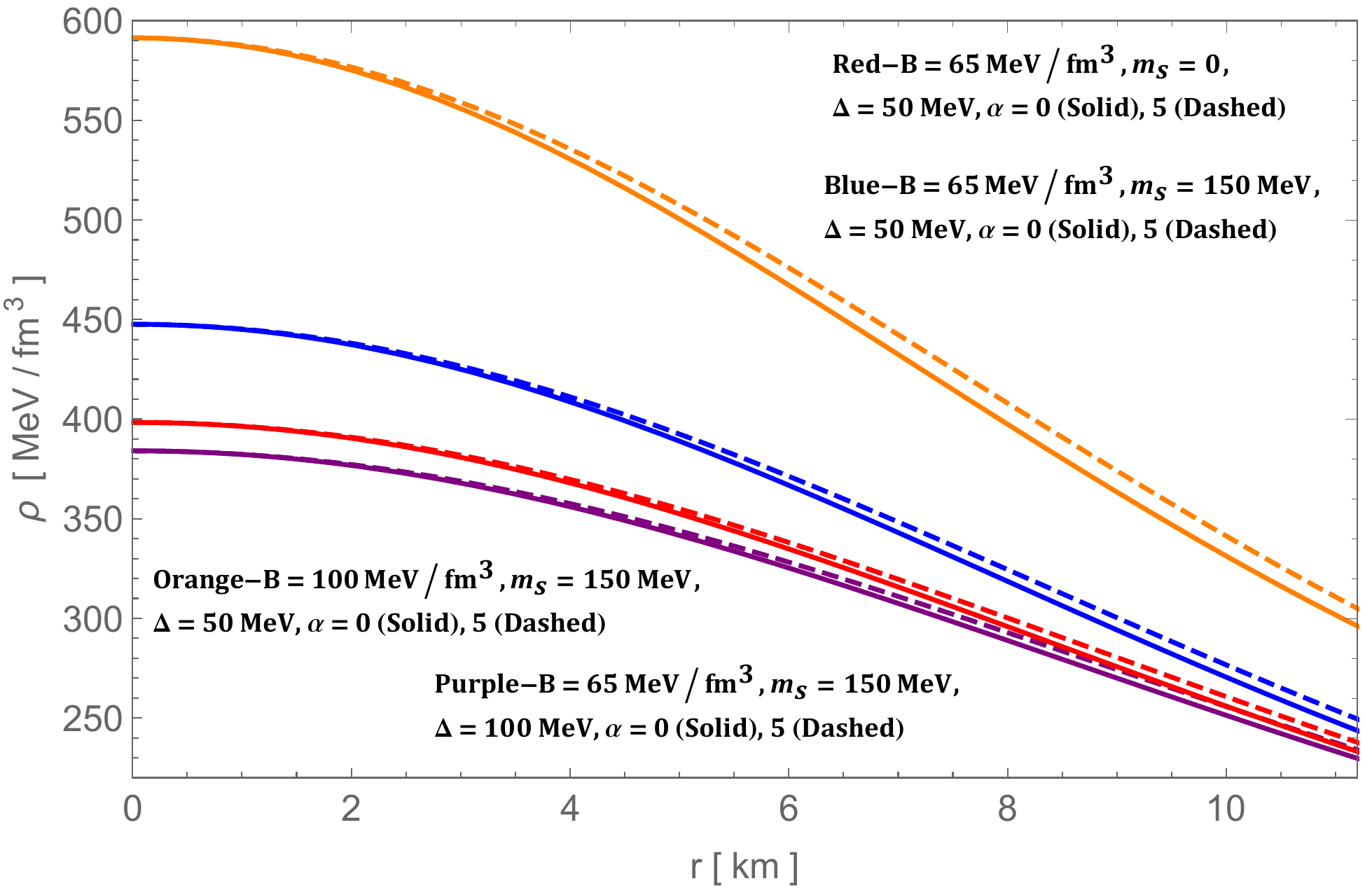}
\caption{Variation of energy density with radius. The numerical values are same as of Fig. \ref{f1}. }
\label{f2}
\end{figure}

\section{Color-flavor locked strange matter} \label{sec3}

Here, we discuss the CFL quark matter in compact stars.
If quark matter is in CFL phase, the thermodynamic potential for  electric and color charge neutral CFL quark matter is given by \cite{alf01} 
\begin{equation}
\Omega_{CFL} = {6 \over \pi^2} \int_0^\nu p^2 (p-\mu)~dp+{3 \over \pi^2} \int_0^\nu p^2 \left(\sqrt{p^2+m_s^2}-\mu \right)~dp-{3\Delta^2 \mu^2 \over \pi^2}+B,
\end{equation}
in order of $\Delta^2$. The symbols have their usual meanings. The first and second terms are contributions from massless $u,~d$ quarks and $m_s$ mass for $s$ quark, while no interaction is considered. The next term is leading correction due to CFL in the power of $\Delta/\mu$ while the final term is the bag constant. At large density regimes, quarks can be considered massless as compared to the chemical potential and also the three-flavor quark is in color-flavor locked state \cite{Alford:1998mk}. The CFL not only trigger the formation of Cooper pairs of different color and flavor but also forced to have same Fermi momentum of all quarks while the electrons are not present \cite{raj01}. The common Fermi momentum with the pairing ansatz in the CFL phase \cite{Steiner:2002gx}, reads
\begin{eqnarray}
n_u = n_d = n_s = {\nu^3+2\Delta^2 \mu \over \pi^2}  ~~~\text{and}~~~
\nu = 2\mu - \sqrt{\mu^2+{m_s^2 \over 3}} \sim \mu-{m_s^2 \over 6\mu},  
\end{eqnarray}
where $n_u$, $n_d$ and $n_s$ are flavor number densities, respectively. The
extra term due to CFL in free energy contributes few percentage of color superconducting gap ($\Delta \sim 0-150 ~MeV$) and the baryon chemical potential ($\mu \sim 300-400 ~MeV$). A MIT based EoS put a strong constraint on the bag constant $B$ to be always greater than 57 $MeV/fm^3$ \cite{far84}. In fact, the free energy contributed from CFL pairing is more than the free energy consumes to maintain equal number of quark densities \cite{alf01}, and therefore CFL paired quarks are more stable than unpaired.\\

The exact expression for an EoS is very difficult to obtained for $m_s \neq 0$. However, for the case of $m_s=0$, one can obtained a simple EoS similar to MIT-bag model with an extra term from CFL contribution as $\rho = 3p+4B-6\Delta^2 \mu^2 / \pi^2$. Further, the exact expression of EoS when $m_s \neq 0$ can be obtained by limiting the series upto $\Delta^2$ and $m_s^2$ terms. In this case the pressure and energy density can be written as \cite{lugo02}
\begin{eqnarray}
p &=& {3\mu^4 \over 4\pi^2}+ {9\beta \mu^2 \over 2\pi^2}-B ~,~~\mbox{and}~~~\rho = {9\mu^4 \over 4\pi^2}+ {9\beta \mu^2 \over 2\pi^2}+B,
\end{eqnarray}
with $\beta = -m_s^2/6+2\Delta^2/3$. Therefore, the total energy density in the mixed phase is
\begin{equation}
\rho = 3p+4B-{9\beta \mu^2 \over \pi^2}~,~~\mbox{provided}~~~\mu^2 = -3\beta + \left[{4\pi^2 (B+p) \over 3}+9\beta^2 \right]^{1/2}.  \label{e3.4}
\end{equation}
The EoS, in the form $P = P(\rho)$, is essentially
determined by the value of the bag constant $B$.

\begin{figure}[h]
\centering
\includegraphics[width=0.6\linewidth]{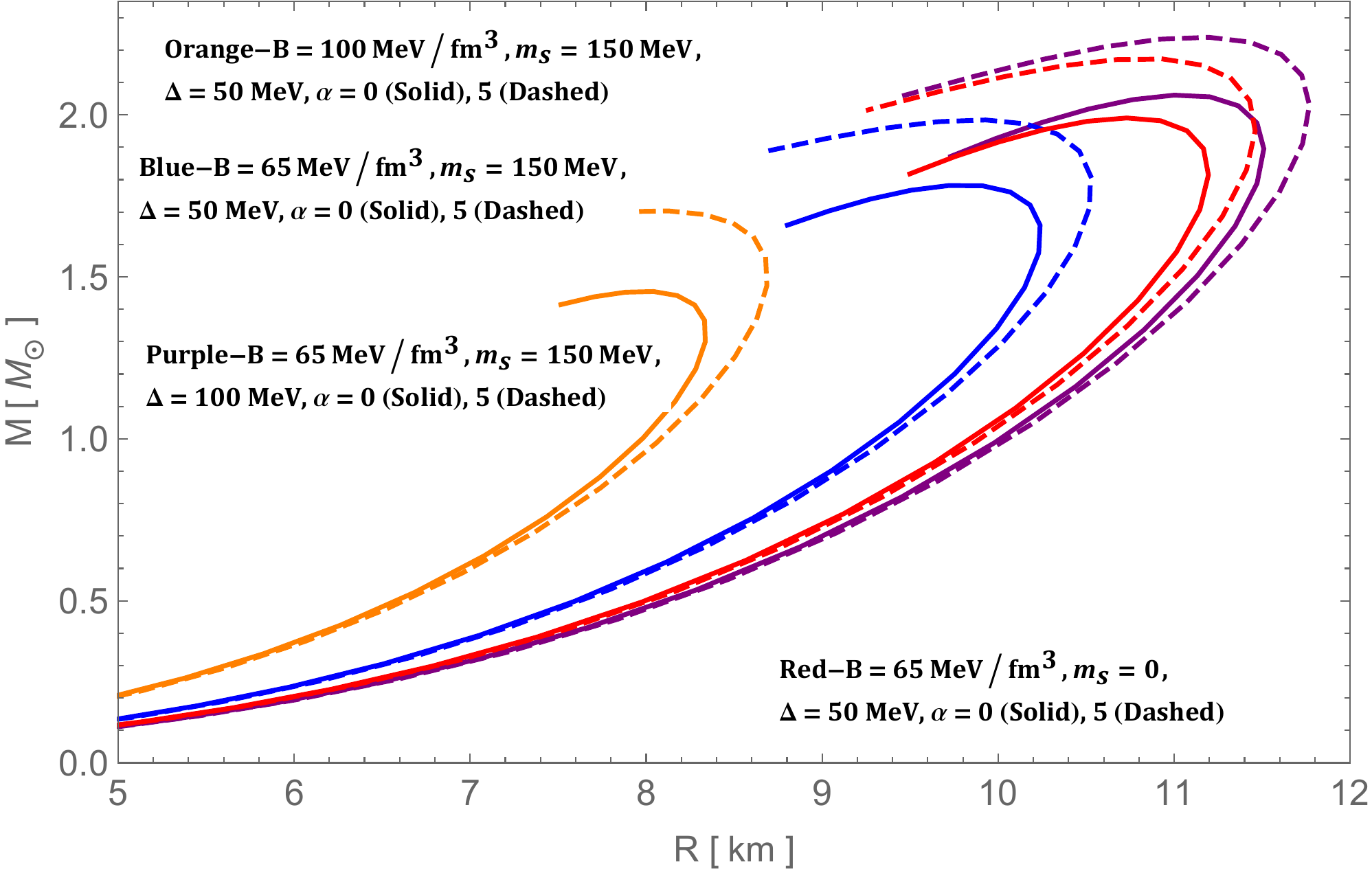}
\caption{Mass radius curves obtained from the of CFL stars as a function of the radius $R$  as in Fig. \ref{f1} with
 two values of $\alpha$. Stellar properties for $\alpha=0$ for GR case and $\alpha >0$ for 4$D$- EGB gravity. }
\label{f3}
\end{figure}

\section{Numerical techniques and results}\label{sec4}
Employing CFL phase at sufficiently high density and relatively low temperature, we have to solve 
three equations with four unknown functions, which are $m(r)$, $\Phi(r)$, $p(r)$ and  $\rho(r)$. Since, the CFL EoS given in \eqref{e3.4}, which is explicitly depends on the bag constant $B$, strange quark mass $m_s$ and color superconducting gap $\Delta$. As a first step, the TOV equation \eqref{e2.11} and mass function \eqref{e2.12} are re-scaled so that the mass is measure in solar mass ($M_\odot$), radius in $km$ whereas density and pressure are in $MeV/fm^3$. Further, the EoS \eqref{e3.4} is re-scaled so that the pressure, density \& $B$ in $MeV/fm^3$ and $m_s$ \& $\Delta$ in $MeV$. In this approach we shall treat the values of $B,~m_s, ~\alpha$ and $\Delta$  as free constant parameters. Given the set of differential equations i.e. \eqref{e2.11} and \eqref{e2.12} with the EoS,  we need an appropriate initial ($\mathcal{I}$) and boundary ($\mathcal{B}$) conditions, which we choose as,
\begin{eqnarray}
\mathcal{I} &:& ~~p(r_0)=p_0 ~~,~~ m(r_0)=\frac{r_0^3}{4 \alpha } \left(\sqrt{\frac{32 \pi \alpha }{3} ~\rho (r_0)+1}-1\right),\\
\mathcal{B} &:& ~~m(R)= M, ~~~ p(R)=0,
\end{eqnarray}
where $p_0$ is central pressure, $R$ is the star radius and $M$ is the total gravitation mass of the star. The procedure in our numerical simulations is discussed below. For solving the equation, we treat the problem as initial-value problem at the center $r_0=0$ and $p_0=53 ~MeV/fm^3$. It is to be noted that the initial pressure (or central pressure) $p_0$ can be freely chosen provided its value lies within the physical limit as predicted by other authors. The pair of defined equations are solved simultaneously for pressure and mass from $r=r_0=0$ to 20 $km$ until the pressure vanishes, which defines the boundary. This leads to the quark star radius $R$ and mass $M = m(R)$.

Thus, one can generate $M-R$ and $p-r$ diagrams. Further, the energy density curves can be generated through the EoS given in \eqref{e3.4}, and rest of the graphs have been drawn using pressure and density. In our case, we fix the constant values as $B=65,~100~MeV/fm^3$, $m_s=0,~150~MeV$, $\Delta=50,~100 ~MeV$ when $\alpha = 0 ~\& ~ 5$. In the limit $\alpha \to 0$, as one can easily see that the solution reduces to GR. Here, we fixed our range of coupling constant $\alpha$ as suggested in Ref. \cite{Doneva:2020ped}. They imposed a strong constraint on $\alpha$ as per the existence of stable stellar black holes that $\sqrt{\alpha} \lesssim 2.6 ~km$ or equivalently $\alpha  \lesssim 6.76~km^2$. Interestingly, we compare our results in 4$D$ EGB gravity as well as in the GR.

For the same central pressure $p_0=53~MeV/fm^3$, in all the cases, as $\alpha$ increases from 0 (solid) to 5 (dashed) the radius of the star as well as the energy density increases which are presented in  Figs. \ref{f1} and \ref{f2}). Also, when the strange quark mass $m_s$ changes from 0 to 150 $MeV$ while the rest are fixed at $B=65~MeV/fm^3,~\Delta=50~MeV$ the radius of the stellar object decreases, Fig. \ref{f1} (Red and Blue). Similarly, for fixed $m_s=150~MeV,~\Delta=5~MeV$ when the bag constant increases from $65~MeV/fm^3$ to $100~MeV/fm^3$ the radius reduces drastically (Fig. \ref{f1}, Blue and Orange). However, when $B=65~MeV/fm^3,~m_s=150~MeV$ and $\Delta$ changes from 50 $MeV$ to 100 $MeV$ the radius increases significantly (Fig. \ref{f1}, Blue and Purple). From Fig. \ref{f2}, one can see that when strange quark mass and bag constant $B$ increases the energy density increases. It was shown that when color-superconducting gap increases the energy density decreases.

\section{Properties of the spheres }\label{sec5}

\subsection{Energy conditions}

In this section, we will briefly study the energy conditions, that are  sets of inequalities depending on energy momentum tensor. To be specific, we start by finding  strange stars for weak energy condition (WEC), i.e. $T_{\mu\nu} U^{\mu}U^{\nu}$, where $U^{\mu}$ is
a timelike vector. For the given diagonal EM tensor, the WEC
implies
\begin{equation}\label{EC1}
\rho (r)\geq 0 ~~\text{and}~~ \rho (r)+ p(r)\geq 0,  
\end{equation}
it follows that if WEC is satisfied then NEC also satisfied. The NEC is the assertion that for any null vector $k^{\mu}$, we should have $T_{\mu\nu} k^{\mu} k^{\nu}\geq 0$. The NEC is the simplest energy condition to deal with algebraically.

\begin{figure*}
\centering
\includegraphics[width=0.6\linewidth]{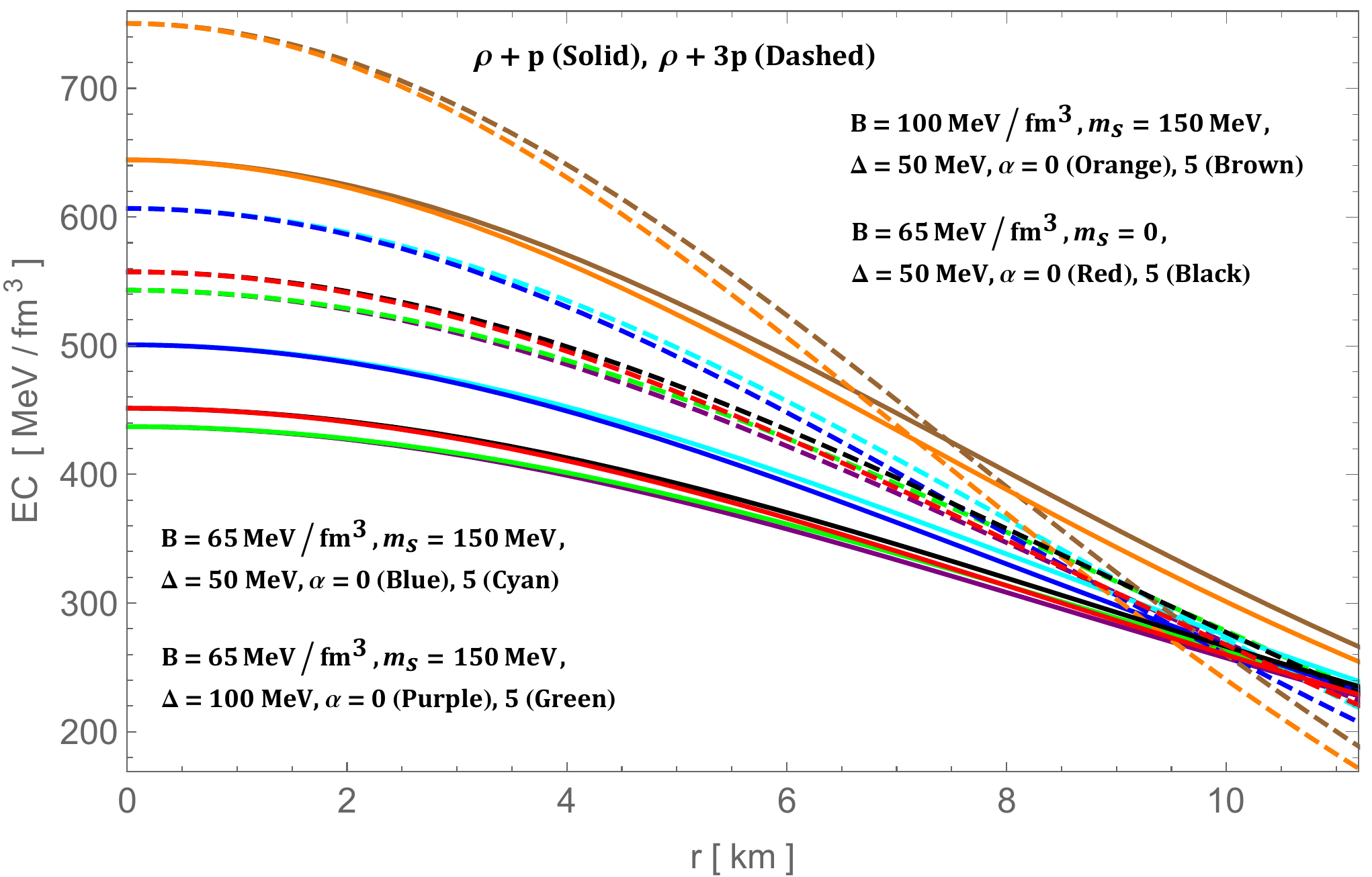}
\caption{Variation of energy conditions $\rho+p$ and $\rho+3p$ with radius for the values given in the caption. The GR value is for $\alpha=0$, while $\alpha >0$ represents $4 D$ EGB gravity. }
\label{f4}
\end{figure*}

Finally, the strong energy condition (SEC) asserts that $\left(T_{\mu\nu}-\frac{1}{2} T g_{\mu\nu}\right) U^{\mu} U^{\nu}\geq 0$ for any  timelike vector $U^{\mu}$.  The strong energy condition (SEC) asserts that gravity is attractive,
\begin{equation}\label{EC3}
\rho (r)+ \sum \mathcal{P}_{i}(r)\geq 0, ~~\text{and}~~ \rho (r)+3 \mathcal{P}(r)\geq 0.
\end{equation}
 Note that the SEC does not imply the WEC, but it follows that any violation of the NEC also violates the SEC and WEC. Considering the above conditions (\ref{EC1}-\ref{EC3}), our results are reported in Fig. \ref{f4}. 
According to Fig. \ref{f4}, we observe that all energy conditions are satisfied. As a result, our 
assumed EoS is suitable for compact quark stars. 

\subsection{Speed of sound within the  fluid and the causality condition}
In order to see more clearly the extremely compact
limit of the objects studied,  we analyse the speed of sound propagation $v^2_s $, which is defined through the equation $v^2_s = dp/ d\rho$. As for the equation of state of the matter, the velocity of sound does less than the velocity of light. Thus, the behavior of the sound speed is always less than unity, as we fix here $c = 1$.

\begin{figure}
\vspace{0.3cm}
    \centering
    \includegraphics[height=4.7cm,angle=00]{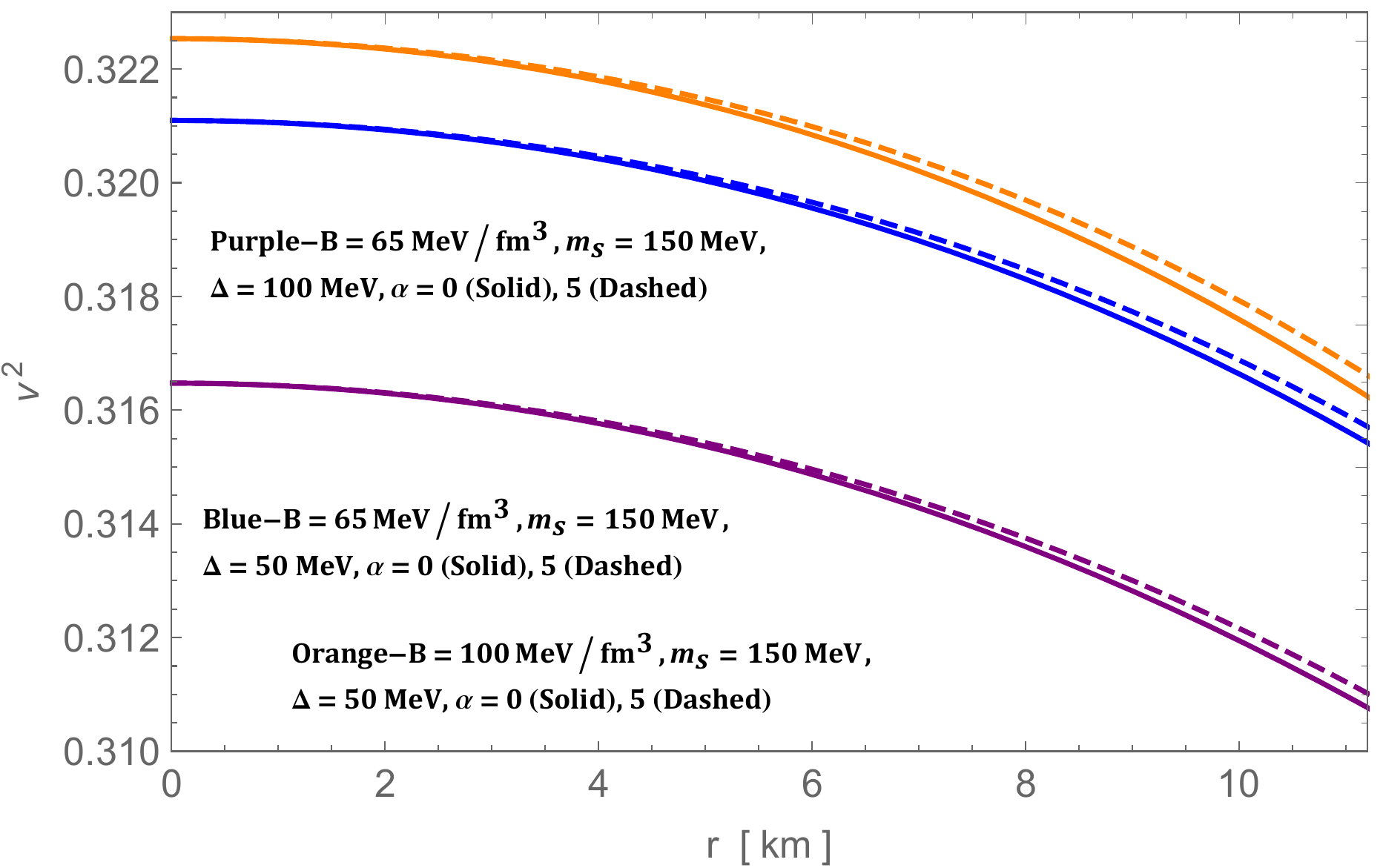}
     \includegraphics[height=4.8cm,angle=00]{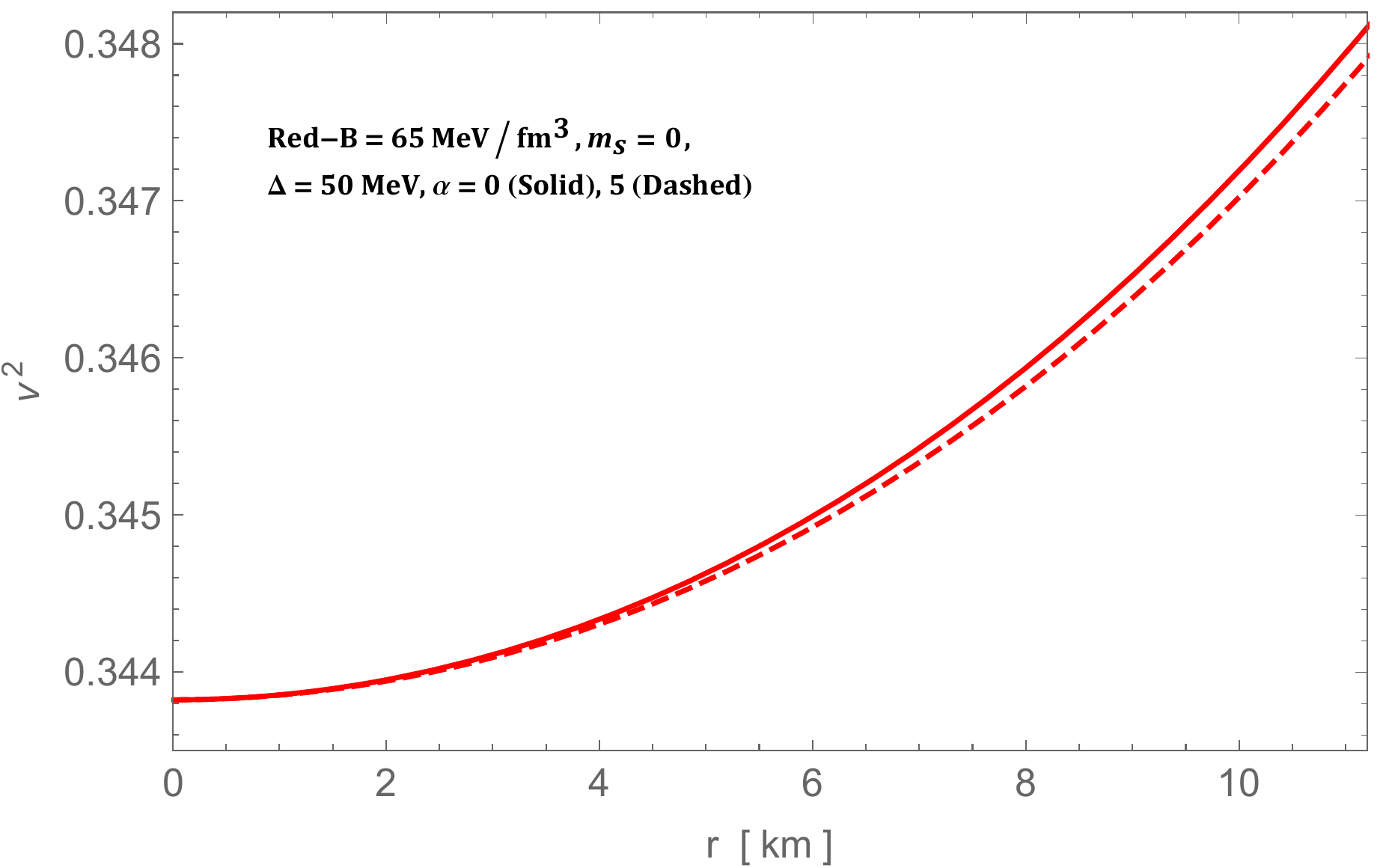}
    \caption{Variation of sound speed with radius for the fixed values given in the caption as well as in sec. \ref{sec4}.}
    \label{f5}
\end{figure}

The causality condition obtained for the four different
cases here considered is shown in Fig. \ref{f5}, together with the solution of the classical relativity ($\alpha$ = 0).
The Fig. \ref{f5} shows the value is maximum when  $(B,~m_s,~\Delta)$ are minimum i.e. $(65 ~MeV/fm^3,~0,~50~MeV)$, and further deceases when $\alpha$ increases from $0 \to 5$ . However, the trend is not the same when we consider $m_s \neq 0 ~(i.e. ~150~MeV)$. For this case, when $\alpha$ and $B$ increases, the speed of sound also increases while the reverse is true when $\Delta$ increases for fixed ($B,~m_s$), see Fig. \ref{f5}.

\subsection{The stability criterion and the adiabatic indices }
More of the stability of compact star model we focus on 
the adiabatic index ($\gamma$) which is an important thermodynamical quantity. It has been pointed out by Chandrasekhar \cite{Chandrasekhar} to solve the instability problem based on the variational method. We now turn to the question of compact stars  whose constituents can be used to build their EoS and consequently their adiabatic index, used to infer their stability. The final expression for the adiabatic index which is
\begin{equation}
\gamma \equiv \left(1+\frac{\rho}{p}\right)\left(\frac{dp}{d\rho}\right)_S,
\end{equation}
where derivation is performing at constant entropy $S$. Moreover, $dp/ d\rho$   is the speed of sound in units of speed of light. Therefore, sound speed is an important quantity related directly with the stiffness of the EoS. For instance, Glass \& Harpaz \cite{Glass} have shown that the adiabatic index should exceed 4/3 in a stable polytropic star by an amount that depends on that ratio $\rho/p$ at the centre of the star. Moreover, this value lies between  2 to 4 in most of the neutron stars equations of state \cite{Haensel}.

\begin{figure}[h]
\centering
\includegraphics[width=0.6\linewidth]{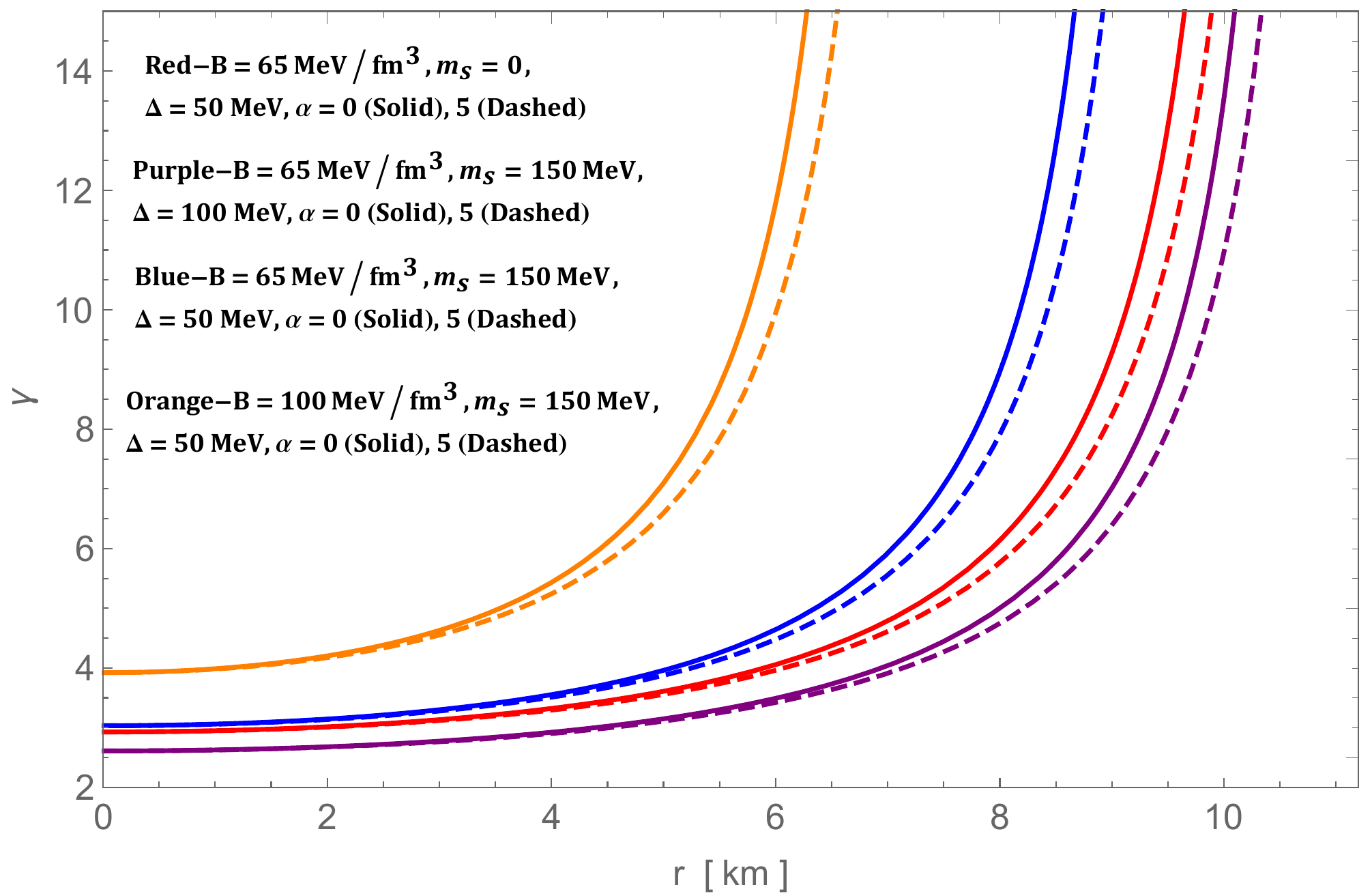}
\caption{Variation of adiabatic index with radius for the values given in the caption.}
\label{f6}
\end{figure}

The stellar structure requires the dynamical stability condition, and $\gamma $  should be more 
than 4/3 $(i.e~ \gamma > 1.33)$ everywhere within the strange star. The results reported in Fig. \ref{f6}  are obtained for values same as Fig. \ref{f1}. Adopting positive $\alpha$, this stellar model is stable  against the radial adiabatic infinitesimal perturbations. 

Stiffness of the EoS for different values of ($B,~m_s,~\Delta$) can we seen for adiabatic index. In Fig. \ref{f6}, one thing is clear that $\gamma$ has the maximum value at $r=0$ for ($100~MeV/fm^3,~150~MeV,$ $~50~MeV$) i.e. for fixed ($m_s,~\Delta$) as $B$ increases the stiffness is also raised. However, the increase in $B$ reduces the boundary because the pressure vanishes rapidly at smaller radii. Therefore, the matter contain inside the interior  can't get much higher resulting into smaller ($R,~M_{max}$) so as the compactness $M/R$ (Fig. \ref{f3}). Moreover, increasing $m_s$ makes the EoS stiffer and softer while $\Delta$ increases. An interesting affect due to coupling constant $\alpha$ is that $\gamma$ remains the same near the core and slightly reduces at the outer layer.

\subsection{$M-R$, $M-I$ curves and stiffness of EoS}

\begin{figure}[h]
\centering
\includegraphics[width=0.6\linewidth]{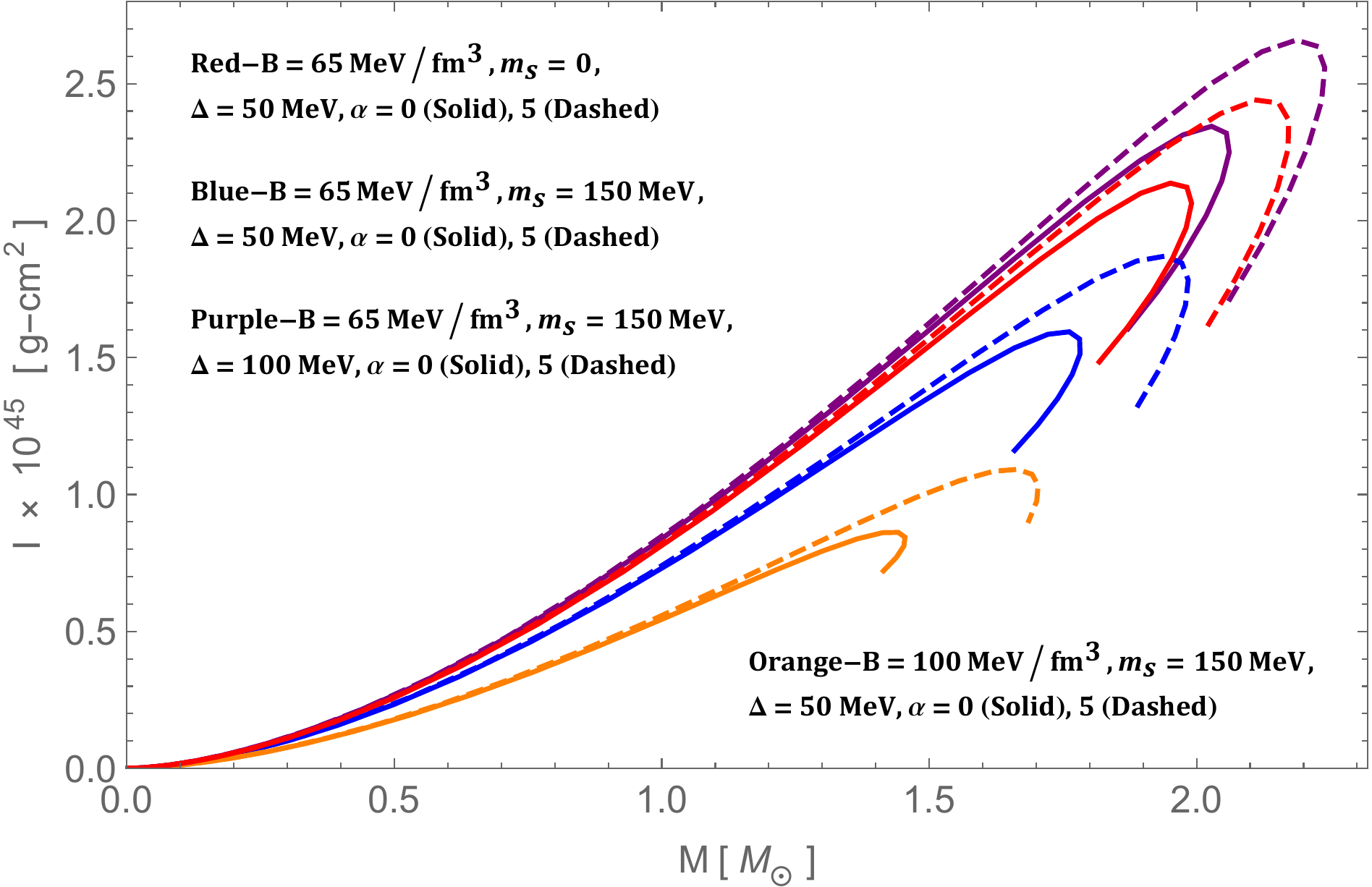}
\caption{$M-I$ curves for the values given inside the frame.}
\label{f7}
\end{figure}

To see the overall stiffness of the CLF EoSs imposing several constraints on ($B,~m_s,~\Delta$) can be seen very clearly from $M-R$ and $M-I$ curves. In Fig. \ref{f3}, the minimum $M_{max}$ is obtained for $(100~MeV/fm^3,~150\,MeV,~50\,MeV)$ (Orange) making it most soft EoS. If $B$ increases from 65 (Blue) to 100 (Orange) $MeV/fm^3$ one can observe a decrease in $M_{max}$ while ($m_s,~\Delta$) are fixed at ($150\,MeV,~50\,MeV$) implying that increasing bag constant $B$ makes EoS soft. On the other hand, if $\Delta$ increases from 50 (Blue) to 100 (Purple) $MeV$ while rest at fixed at ($65\,MeV/fm^3,~150\,MeV$) there is a sharp increase in $M_{max}$ i.e. stiffness increases with color-superconducting gap $\Delta$. At last, while the strange quark mass increase from 0 (Red) to 150 (Blue) $MeV$ keeping ($B,~\Delta$) fixed at ($65\,MeV/fm^3,~50\,MeV$) the $M_{max}$ again reduces i.e. increasing $m_s$ makes the EoS softer. Similar affects can also be observed from $M-I$ curve, see Fig. \ref{f7}. Further, the EGB modification in the geometric sector strongly influence the stiffness of the CFL EoS. In fact, $M-R$ curves (Dashed) generated in 4$D$ EGB theroy can have more $M_{max}$ than the GR counterparts (Solid), that can be seen in Fig. \ref{f3}.\\

Having derived the $M-I$ curves that describe slow rotation approximation, we use Bejger \& Haensel \cite{bej} formula, which is
\begin{equation}
I = {2 \over 5} \left(1+{M \over R} \cdot {km \over M_\odot} \right) MR^2. \label{e5.4}
\end{equation}
Since, Eq. \eqref{e5.4} enable a static configuration without considering rotation into the interior spacetime to determine the moment of inertia which is accurate within 5\%. Incorporating $M-R$ variations in \eqref{e5.4} generates the required $M-I$ curve.

\section{Summary and Discussion }\label{sec6}
In this work we have performed a detailed study on the properties of color-flavor locked quark matter in compact star interior. This paper presents a systematic  study of static, spherically symmetric solutions in 4$D$ Einstein-Gauss-Bonnet gravity, which bypasses the conclusions of Lovelock's theorem and avoids Ostrogradsky instability. The hydrostatic equilibrium equations are obtained in order to test the new theory in strange stars whose mass-radius diagrams are obtained using CFL quark matter and compared with known GR solutions.

The high density and relatively low temperature required to produce color superconducting quark matter may be attained in quark stars. This opens up the possibility of using astronomical observations to identify hybrid neutron stars or strange stars.  Here, using an EoS derived within the MIT bag model, we solve the field equations numerically and obtain the $M-R$ curves for CFL quark stars. 

In order to solve satisfactorily the mass function and TOV equations, we establish a range for $\alpha >0$
as the expression is always gives positive solution. Since, the existence of stellar mass black holes can constrain the values of $\sqrt{\alpha} \lesssim 2.6~km$, when $\alpha >0$. We restrict our discussion to the positive branch of solution.  An interesting aspect related to our results is the fact that increases the value of $\alpha$,  clearly controls the values of the maximum star masses. We have established that  increases the value of $\alpha$, the contribution coming from GB term, one can obtain greater values for the masses of the CFL phase in compact stars. The lower limit is obtained in such a way that at least a $1.44~ M_{\odot}$ star can be achieved for GR case when the maximum value of $B=100~MeV/fm^3$ (see Fig. \ref{f3}). One thing is very clear from the current investigation that while increasing bag constant and strange qurak mass, the stiffness of CFL EoS reduce thereby reducing $M_{max}$ while increasing color-superconducting gap and EBG coupling term makes it stiffer. It is also found that the behavior of the  equilibrium compact stars is very similar to the GR case, which was also reported in \cite{Doneva:2020ped}. Moreover, we make some remarks about the general physical properties of CFL strange stars. We have found through our numerical scheme that velocity of sound, energy conditions and  adiabatic stability are in favour of the physical requirements, as we can see from Figs. \ref{f4}-\ref{f7}.

Finally, we make some remarks about the CFL strange star in the 4$D$ EGB gravity. If we accept the quark matter is in CFL state, we can explain the other physical properties of 
strange stars with other choices of EoS, but we expect that our numerical results do not change drastically.

\end{document}